\documentclass{emulateapj}

\pdfoutput=1
\usepackage{epstopdf}
\usepackage{amsmath}
\usepackage{comment} 
\usepackage{appendix}
\usepackage{enumerate}
\usepackage{natbib}
\usepackage{color}

\slugcomment{\today}

\newcommand{\sigMINUS}{\sigma_{-}}
\newcommand{\sigPLUS}{\sigma_{+}}

\begin{document}

\newcommand{\gps}{\ensuremath{g_{\rm p1}}}
\newcommand{\rps}{\ensuremath{r_{\rm p1}}}
\newcommand{\ips}{\ensuremath{i_{\rm p1}}}
\newcommand{\zps}{\ensuremath{z_{\rm p1}}}
\newcommand{\yps}{\ensuremath{y_{\rm p1}}}
\newcommand{\wps}{\ensuremath{w_{\rm p1}}}
\newcommand{\griz}{\gps\rps\ips\zps}
\newcommand{\PS}{\protect \hbox {Pan-STARRS1}}

\title{Measuring Type Ia Supernova Populations of Stretch and Color
 and Predicting Distance Biases}

\shorttitle{Distance Biases}
\shortauthors{Scolnic \& Kessler}

\def\uch{1}

\author{
D. Scolnic\altaffilmark{\uch}\&
R. Kessler\altaffilmark{\uch}
}

\altaffiltext{\uch}{Department of Physics, The University of Chicago, Chicago,IL 60637, USA.  Email: dscolnic@kicp.uchicago.edu, kessler@kicp.uchicago.edu}

\begin{abstract}
Simulations of Type Ia Supernovae (SN\,Ia) surveys are a critical tool for correcting biases in the analysis of SN\,Ia to infer cosmological parameters.  Large scale Monte Carlo simulations include a thorough treatment of observation history, measurement noise, intrinsic scatter models and selection effects.  In this paper, we improve simulations with a robust technique to evaluate the underlying populations of SN\,Ia color and stretch that correlate with luminosity.  In typical analyses, the standardized SNIa brightness is determined from linear `Tripp' relations between the light curve color and luminosity and between stretch and luminosity.  However, this solution produces Hubble residual biases because intrinsic scatter and measurement noise result in measured color and stretch values that do not follow the Tripp relation.  We find a $10\sigma$ bias (up to 0.3 mag) in Hubble residuals versus color and $5\sigma$ bias (up to 0.2 mag) in Hubble residuals versus stretch in a joint sample of 920 spectroscopically confirmed SN\,Ia from PS1, SNLS, SDSS and several low-z surveys.  After we determine the underlying color and stretch distributions, we use simulations to predict and correct the biases in the data.  We show that removing these biases has a small impact on the low-z sample, but reduces the intrinsic scatter $\sigma_{\textrm{int}}$ from $0.101$ to $0.083$ in the combined PS1, SNLS and SDSS sample.  
Past estimates of the underlying populations were too broad, leading to a small bias in the equation-of-state of dark energy $w$ of $\Delta w=0.005$.  
 \end{abstract}

\section{Introduction}
\label{sec:intro}

The standardization of measurements of Type Ia Supernovae led to the discovery that the universe is expanding at an increasing rate (\citealp{Riess98}, \citealp{Perlmutter99}).  The cause of this acceleration, dubbed `dark energy', is one of the great mysteries in cosmology today.  Because SN\,Ia remain one of the premier tools to measure dark energy properties, great effort has gone into acquiring increasingly large samples and refining methods of standardization.

Conventionally, the diversity of SN\,Ia light curves is described by three parameters: an amplitude, a color and a light curve width.  The Tripp estimator \citep{Tripp} uses these three parameters to standardize the SN\,Ia brightness and determine the distance.  Within the formalism of the light curve model SALT2 \citep{Guy10}, the distance modulus, $\mu$, for each SN\,Ia is expressed as
\begin{equation}
\mu=m_b+\alpha x_1-\beta c-M,
\end{equation}
where $m_b$ is the log of the overall amplitude needed to scale the model to match the data, $x_1$ describes the width or `stretch' of the light curve, $c$ is a color parameter that is roughly $B-V$ at peak and is more positive for more red colors, and $M$ is the rest-frame magnitude of an SN\,Ia with $\alpha=\beta=0$.  The global nuisance parameters $\alpha$ and $\beta$ quantify the correlation between luminosity with stretch and color, respectively. 

The Tripp estimator does not capture all of the brightness variation of SN\,Ia, and recent simulation studies (\citealp{Kessler2013} (K13), \citealp{Mosher14}, \citealp{Scolnic14a} (S14a), \citealp{S14b} (S14b))  have shown there are small biases in the measurements of $\alpha$ and $\beta$.  The fundamental issue with the Tripp estimator is that it does not separate the amount of color or stretch that correlates with luminosity, here referred to as the underlying color or stretch population, from intrinsic scatter of the SN brightness or measurement noise.  The intrinsic scatter is the magnitude of the dispersion of distance residuals to a best fit cosmology if light curves were observed with no measurement noise.  This scatter can be caused by either chromatic or achromatic variation of the spectral model.  Near the magnitude limit of a survey, intrinsic scatter and measurement noise result in preferentially selecting events that are brighter than the standardized brightness based on the measured color and stretch. This selection bias results in a $\mu$ bias that increases with redshift, and if not corrected, results in a significant bias in the inferred cosmological parameters. Additional bias occurs when fluctuations result in color and stretch values that do not correlate with luminosity, but are still assumed to follow Eq. 1.  For example, a `true' SN color of $c=0$ may be observed at $c=-0.2$ due to intrinsic scatter and measurement noise and there will then be a distance bias of $-\beta\times(c_{\textrm{Obs}}-c_{\textrm{True}}) \sim 0.6$ mag as $\beta\sim 3$.  Previous analyses (e.g. \citealp{Jha07}, \citealp{Mandel11}) have used Bayesian models to separate the components of the observed color that are due to reddening, intrinsic scatter and noise.  These approaches assume a prior on the component of color that correlates with luminosity and this prior is described by a single parameter exponential function.  However, these approaches do not account for survey selection effects or the stretch population.  Our analysis robustly determines a general parameterization of the components of both color and stretch that correlate with luminosity, and we include a full model of selection effects into our treatment.

The analysis discussed in this paper uses a large sample of $920$ spectroscopically confirmed SN\,Ia; data is combined from CfA1-4 (\citealp{Riess99}, \citealp{Jha06}, \citealp{Hicken09a}, \citealp{Hicken12}), CSP \citep{Stritzinger11}, PS1 \citep{Rest14}, SDSS \citep{Sako14} and SNLS \citep{Guy10}.  An overview of the calibration of these samples is presented in \cite{Supercal}.   Previous analyses (K13, S14a, \citealp{Betoule14} (B14)) have determined biases from Eq. 1, but the underlying color and stretch distributions were not rigorously determined.  These studies all used simulations of SN surveys generated by the SNANA software package \citep{SNANA}.  These simulations have been extensively tested where they show excellent agreement with the data in several key distributions.  The simulations include models of the intrinsic scatter, measurement noise and selection effects.  The data and simulation input files used in this analysis are publicly available in the SNANA package \footnote{See SNDATA\_ROOT/sample\_input\_files/SK16}.
        
In section 2, we describe the formalism to determine the underlying color and stretch distributions.  In section 3, we predict the biases in Hubble residuals and correct for them.  In section 4, we present our discussion and conclusions.

\section{Determining the underlying color and stretch populations}

We follow the formalism presented in K13 in which an underlying color population is parameterized by three parameters describing an asymmetric gaussian such that 
\begin{equation}
P(c)=
\left\{
\begin{array}{ll}
 e^{ [-(c - \bar{c})^2/2\sigMINUS^2] } &  \mbox{if } c < \bar{c};\\
  e^{ [-(c - \bar{c})^2/2\sigPLUS^2] } & \mbox{if } c > \bar{c},~  
\end{array}
\right.
  \end{equation}
where $\bar{c}$ is the value with maximum probability and $\sigMINUS$ and $\sigPLUS$ are the low and high sided gaussian widths.  A similar parameterization describes the stretch $x_1$.  This parameterization is chosen to account for the observed asymmetric distributions that have more events for dimmer SNe which are redder and have smaller stretch.  To determine the 3 color and 3 stretch parameters describing the asymmetric gaussians, we find the set of parameters for which the simulations of the PS1, SDSS, SNLS and low-z surveys best match the data from these surveys.  While matching the color and stretch distributions could be done over a grid of asymmetric gaussian parameters, we have instead developed a much faster fitting method that requires just a single simulated sample.  

To measure the underlying populations, we simulate a flat distribution of both color and stretch, and measure the migration of input values to observed values.  The measurement noise and intrinsic scatter smear the input distributions, and selection effects skew the observed distribution towards brighter SNe with bluer colors and higher-stretch values. We define a migration matrix where each component $C_{ij}$ is the ratio of simulated SNe with a fitted color $c_i$ to the number of simulated SNe with a true input color $c_j$.   A similar migration matrix with $X_{ij}$ describes the stretch parameter.  Without measurement noise, intrinsic scatter or selection effects, these matrices are purely diagonal.  Adding measurement noise and intrinsic scatter, the matrix has non-zero off-diagonal elements.  Adding selection effects, the migration matrices are skewed towards bluer colors and higher stretch values.  In this analysis, we analyze the color and stretch populations independently.  This is likely not fully adequate if there are correlations between the distributions of the color and stretch \citep{Sullivan10}.  However, initial tests show that the reduced covariance between $x_1$ and $c$ is $<5\%$.

To solve for the underlying color population, $P(c)$, we first define a binned distribution $\vec{P_c}$ where $P_{cn}$ is the probability that SNe have a color in bin $n$.  We then define a data-simulation difference vector $\vec{\Delta_c}$:
\begin{equation}
\begin{aligned}
\vec{\Delta_c}= 
\Bigg(
\begin{bmatrix}
    o_{c1}  \\
    o_{c2} \\
    \vdots \\
    o_{cn}
\end{bmatrix}
-
\begin{bmatrix}
    c_{11} & c_{12}  & \dots  & c_{1n} \\
    c_{21} & c_{22}  & \dots  & c_{2n} \\
    \vdots & \vdots  & \ddots & \vdots \\
    c_{d1} & c_{d2}  & \dots  & c_{dn}
\end{bmatrix}
\times
\begin{bmatrix}
    P_{c1}  \\
    P_{c2} \\
    \vdots \\
    P_{cn}
\end{bmatrix}
\Bigg)
\end{aligned}
\end{equation}where the binned vector $o_c$ represents the observed distribution of fitted color with the same binning as the migration matrix $C$ and the population vector $\vec{P_c}$.  A similar equation is used for the underlying stretch distribution with efficiency matrix $X$, population vector $\vec{P_x}$ and observed vector $\vec{o_{x1}}$.  The population vector is defined using Eq. 2 and only has three independent elements so that it is well constrained.  We also define a data error vector $\vec{e_c}=[e_{c1},e_{c2}...,e_{cn}]$.  To determine the asymmetric gaussian parameters that describe $\vec{P_c}$, we minimize 
\begin{equation}
\chi^2_c=\sum_{i=1}^n \bigg(\frac{{\Delta}_{ci}}{{e}_{ci}}\bigg)^2.
\end{equation}
where ${\Delta}_{ci}$ and $e_{ci}$ are the i'th component in vectors $\vec{\Delta_c}$ and $\vec{e_c}$ respectively.  The error vector is set to be $\sqrt{\vec{o_c}}$ when $\vec{o_c}\neq 0$ and $\vec{e_{c}}=1.0$ when $\vec{o_c}=0$.  Setting $\vec{e_{c}}=1.0$ for empty bins is not technically correct given a purely Poissonian distribution \citep{Baker}.  However, the distribution is not exactly Poissonian and our approximation limits the bias from empty bins that alters the gaussian parameters by $\sim0.2\sigma$ for our combined sample. 

We do not modify any part of the simulations used for the cosmology analyses besides the underlying parameter populations.   The choice of using the spectroscopic selection functions already determined for each survey is slightly inaccurate because each survey selection function was determined after assuming an approximate underlying color and stretch population.  However, varying the selection functions within their uncertainties only changes the underlying population parameters by $<0.1\sigma$ for all the surveys analyzed here.  Similarly, the input cosmological parameters in the simulation, varied within their uncertainties, have $<0.05\sigma$ effect on the recovered populations. 

We limit this analysis to two models of intrinsic scatter with different amounts of chromatic and achromatic spectral variation: 1) the model presented in \cite{Guy10} (hereafter called the G10 scatter model) in which $70\%$ of the contribution to the Hubble residuals comes from achromatic variation and $30\%$ comes from chromatic variation and 2) the model presented in \cite{Chotard11} (hereafter called the C11 scatter model), in which $25\%$ of the contribution to the Hubble residuals comes from achromatic variation and $75\%$ comes from chromatic variation.  Since the G10 and C11 scatter models were originally created as broadband variations models, K13 converted each of these models into a spectral-variation model that is added to the SALT2 spectral model in our simulations.

\begin{figure}
\centering
\epsscale{1.15}  
\plotone{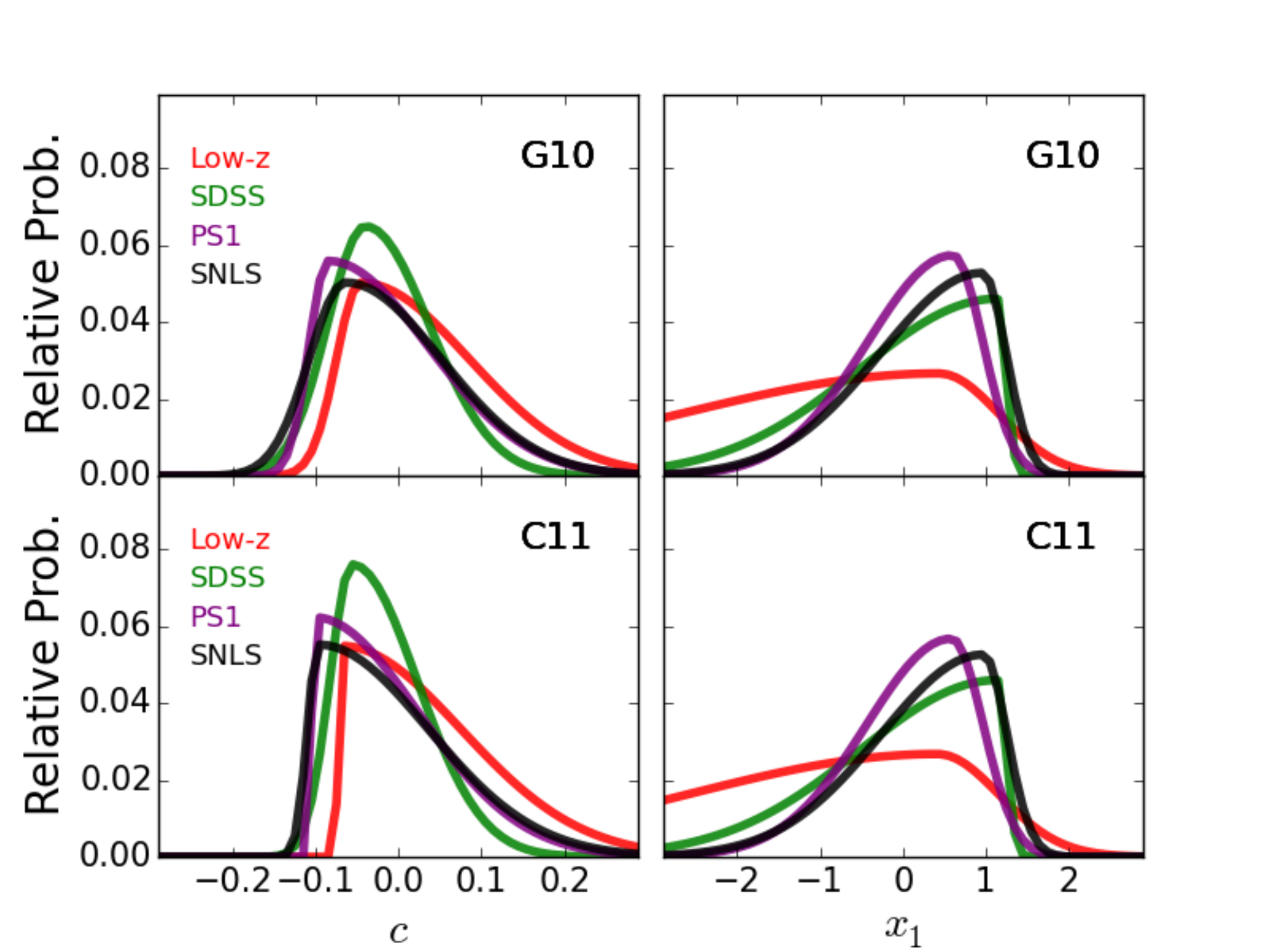}
\caption{The underlying distributions of $c$ (left) and $x_1$ (right) that correlate with luminosity.  A distribution for the Low-z, SDSS, PS1 and SNLS samples are shown, both for the G10 scatter model and the C11 scatter model. }
\label{fig:filters}
\end{figure}

We present our results for underlying color and stretch populations in Table 1 and Fig. 1.   In our analysis, we determine a solution for each survey.  While we do not explicitly solve for redshift-dependent population parameters, the different redshift range for each survey provides some information about the redshift dependence.  The median redshifts for the low-z, SDSS, PS1 and SNLS samples are $z=0.03, 0.20, 0.29, 0.64$ respectively.  We find reasonable consistency in the derived populations among the SDSS, SNLS and PS1 samples, while the low-z population is somewhat different. The different low-z population is expected because low-z surveys typically targeted bright galaxies and thus favored galaxies that host fainter SNe with redder colors and lower stretch values \citep{Sullivan10}.  We also find that the different amounts of chromatic spectral variation have a significant effect on the measured underlying color population, but very little impact on the measured underlying stretch population.  For the C11 model with a large amount of color scatter, the underlying population of color resembles an exponential decay.  This distribution is consistent with extinction by dust being the primary component of the color-luminosity correlation (\citealp{Jha07}, \citealp{Hatano98}).   Most of the underlying distributions show steep drops, particularly at the high-stretch end and the blue-color end.   The $\sigma_{-}$ for $c$ and $\sigma_{+}$ for $x_1$ values that are within 1 standard deviation from 0 indicate a steep drop.

\begin{table}[ht]
\caption{Asymmetric Gaussian Parameters to Describe Underlying Populations.}
  \begin{center}
    \begin{tabular}{ll | c c c }
Survey &Var. & $\bar{c}$ & $\sigMINUS$  &$\sigPLUS$   \\
    \hline
Low-z & G10 &$ -0.055 \pm 0.036 $&$ 0.023 \pm 0.025 $&$  0.150 \pm 0.034$\\
SDSS & G10 &$ -0.038 \pm 0.014 $&$ 0.048 \pm 0.009 $&$  0.079 \pm 0.012$\\
PS1 & G10 &$ -0.077 \pm 0.023 $&$ 0.029 \pm 0.016 $&$  0.121 \pm 0.019$\\
SNLS & G10 &$ -0.065 \pm 0.019 $&$ 0.044 \pm 0.013 $&$  0.120 \pm 0.019$\\
All & G10 &$ -0.043 \pm 0.010 $&$ 0.052 \pm 0.006 $&$  0.107 \pm 0.010$\\
High-z & G10 &$ -0.054 \pm 0.011 $&$ 0.043 \pm 0.007 $&$  0.101 \pm 0.009$\\
Low-z & C11 &$ -0.069 \pm 0.008 $&$ 0.003 \pm 0.003 $&$  0.148 \pm 0.024$\\
SDSS & C11 &$ -0.061 \pm 0.027 $&$ 0.023 \pm 0.020 $&$  0.083 \pm 0.018$\\
PS1 & C11 &$ -0.103 \pm 0.003 $&$ 0.003 \pm 0.003 $&$  0.129 \pm 0.014$\\
SNLS & C11 &$ -0.112 \pm 0.004 $&$ 0.003 \pm 0.003 $&$  0.144 \pm 0.017$\\
High-z & C11 &$ -0.099 \pm 0.003 $&$ 0.003 \pm 0.003 $&$  0.119 \pm 0.007$\\
All & C11 &$ -0.062 \pm 0.016 $&$ 0.032 \pm 0.011 $&$  0.113 \pm 0.014$\\
\hline

~ & ~ & ~ & ~ & ~ \\

    \hline
~  &~& $\bar{x_1}$ & $\sigMINUS$  &$\sigPLUS$   \\
    \hline

Low-z & G10 &$ 0.436 \pm 0.563 $&$ 3.118 \pm 1.582 $&$  0.724 \pm 0.351$\\
SDSS & G10 &$ 1.141 \pm 0.032 $&$ 1.653 \pm 0.076 $&$  0.100 \pm 0.100$\\
PS1 & G10 &$ 0.604 \pm 0.183 $&$ 1.029 \pm 0.138 $&$  0.363 \pm 0.121$\\
SNLS & G10 &$ 0.964 \pm 0.136 $&$ 1.232 \pm 0.098 $&$  0.282 \pm 0.094$\\
High-z & G10 &$ 0.973 \pm 0.105 $&$ 1.472 \pm 0.080 $&$  0.222 \pm 0.076$\\
All & G10 &$ 0.945 \pm 0.100 $&$ 1.553 \pm 0.117 $&$  0.257 \pm 0.078$\\
Low-z & C11 &$ 0.419 \pm 0.559 $&$ 3.024 \pm 1.468 $&$  0.742 \pm 0.347$\\
SDSS & C11 &$ 1.142 \pm 0.455 $&$ 1.652 \pm 0.232 $&$  0.104 \pm 0.305$\\
PS1 & C11 &$ 0.589 \pm 0.179 $&$ 1.026 \pm 0.137 $&$  0.381 \pm 0.117$\\
SNLS & C11 &$ 0.974 \pm 0.128 $&$ 1.236 \pm 0.094 $&$  0.283 \pm 0.088$\\
High-z & C11 &$ 0.964 \pm 0.105 $&$ 1.467 \pm 0.080 $&$  0.235 \pm 0.075$\\
All & C11 &$ 0.938 \pm 0.101 $&$ 1.551 \pm 0.118 $&$  0.269 \pm 0.078$\\
\hline

\end{tabular}
  \end{center}
{Notes: The parameters for each subsample, as well as the High-z compilation (SDSS,SNLS,PS1) and the full compilation (All) are shown.  The first column shows the sample and second column shows the scatter model used in the simulation.  The first part of the table shows the recovered values of the color $c$ population and the second part of the table shows the recovered values of the stretch $x_1$ distribution.}
\end{table}

\begin{figure}
\centering
\epsscale{1.20}  
\plotone{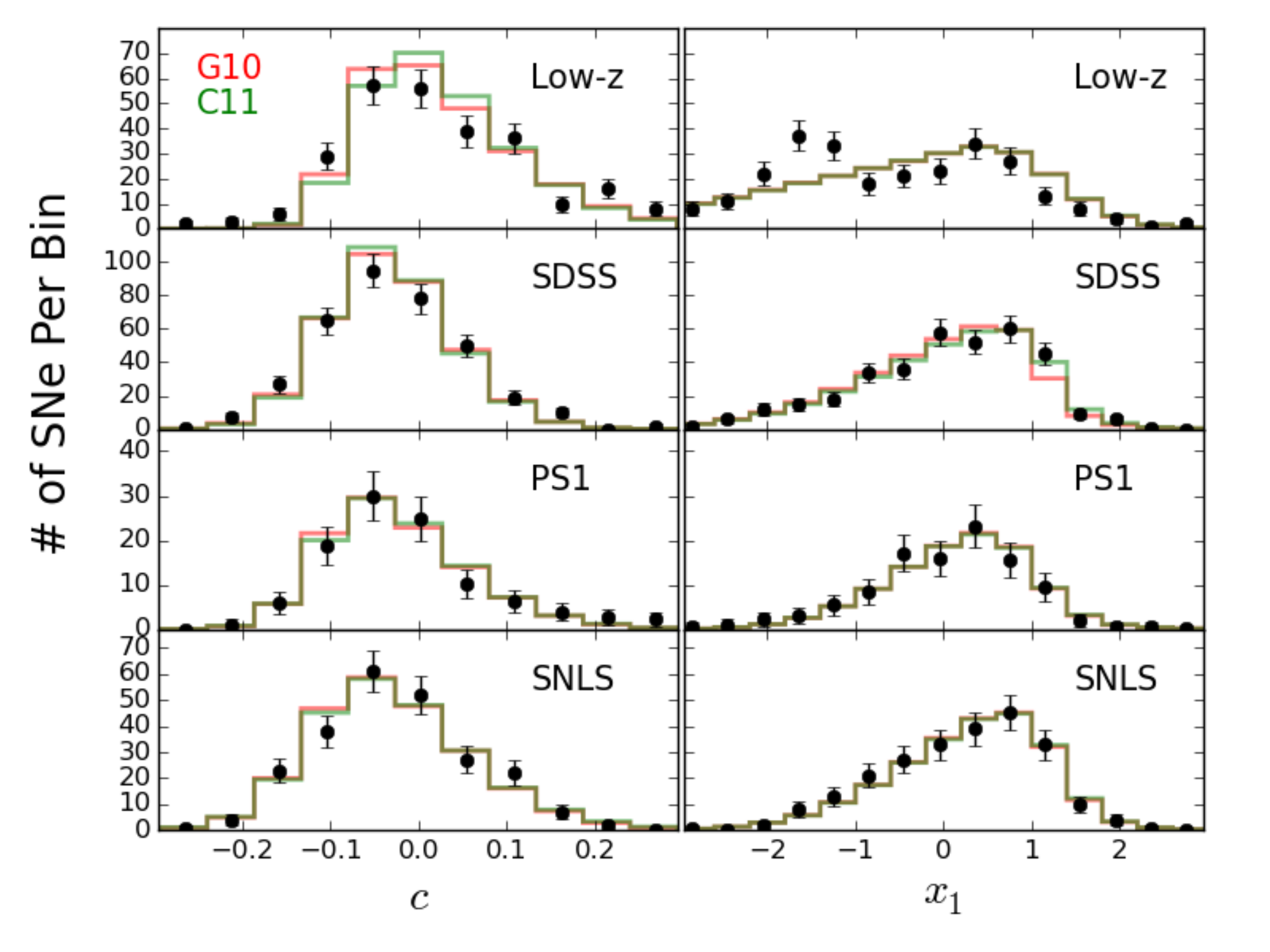}
\caption{The simulated distributions (red, green) of $x_1$ and $c$ using the underlying populations as well as the data (black).  The distributions for each subsample are shown, and the simulated distributions for both the G10 and C11 scatter models are shown.  The size of the simulated distributions is normalized to have the same number of SNe as in the data samples.} 
\label{fig:filters}
\end{figure}

Once we have determined our underlying populations, we simulate samples with different values of $\alpha$ and $\beta$ to see what reproduces the data and estimate biases in the fitted $\alpha$ and $\beta$. From our data compilation, we find $\alpha_{\textrm{Obs}} = 0.142 \pm 0.005$ and $\beta_{\textrm{Obs}} = 3.123 \pm 0.060$. We simulate a grid of Monte Carlo simulations with different input values of $\alpha$ and $\beta$ ($\Delta \beta =0.05, \Delta \alpha=0.005, \beta$ in [2,5] and $\alpha$ in [0.1,0.17]), and then use the SALT2mu program \cite{Marriner11} to fit for $\alpha$ and $\beta$.  We find that simulation inputs of $\alpha_{\textrm{Sim}}$ and $\beta_{\textrm{Sim}}=3.1$ reproduces the data with the G10 scatter model, and inputs of $\alpha_{\textrm{Sim}}=0.14$ and $\beta_{\textrm{Sim}}=3.85$ reproduces the data with the C11 model.

In Fig. 2, we show that when we resimulate our samples with our determined populations, we accurately reproduce the observed distributions.  Of the eight comparisons shown in the figure, only the low-z $x_1$ distribution shows a potential discrepancy between the data and simulations.  For this specific comparison, there are two bins around $x_1=-2$, one with a $3\sigma$ discrepancy and one with $2\sigma$ discrepancy from the simulation, though it is unclear if this is a statistical fluctuation. Comparing Fig. 1 and Fig. 2 illustrates a significant part of the $\mu$ bias. The underlying population drops to zero around $c\sim-0.15$ or $x_1\sim1.5$ (Fig. 1), yet these bins are populated in the measured distributions (Fig. 2).  Therefore, treating these biased color and stretch values with Eq. 1 results in biased distances.  

Once we have determined our underlying populations,

Here we compare our findings for the underlying populations with the JLA (B14) and PS1 (S14b) analyses.  The parameters found for the JLA analysis and PS1 analysis were both based on visual matching of data and simulations.  The JLA analysis overestimates the width of the underlying stretch and color populations for each subsample.  For example, for the SDSS population, JLA uses ($\bar{c}$, $\sigMINUS$,$\sigPLUS$) of (0.0,0.08,0.13) whereas we find ($-0.038$,0.048,0.079).  Furthermore, JLA assumed the same underlying population for both the G10 and C11 scatter models, whereas we find that the color population depends on the scatter model chosen ($\Delta \bar{c}\sim0.02$ for different scatter models).  The parameters found in the PS1 analysis for the G10 scatter model are in good agreement with those found here, but those found for the low-z sample with the C11 scatter model have a significant mean color offset $\Delta c=0.035$ compared to that found here.  

We also compare our measurements of $\beta$ with these previous analyses.  The value of $\beta$ found with the G10 model is similar to many past analyses (e.g. B14).  The value of $\beta$ found with the C11 model agrees with that found in \cite{Chotard11} as well as a recent analysis that examines supernova spectral twins \citep{Fakhouri15}.  

\section{Correcting Hubble biases}

Since both the color and stretch populations have a sharp roll off at the bright end, statistical fluctuations can result in measured $c$ and $x_1$ values beyond the range that correlates with luminosity. We therefore expect Hubble residual biases at the extreme color and stretch values, and here we predict these biases using simulations that include  the underlying populations.  We show trends in Hubble residuals with color and stretch in Fig. 3 for both data and simulations.  Here again, we perform a separate simulation for the G10 and C11 intrinsic scatter model.  We find that both simulations reproduce the significant biases seen in the data near bluer colors and higher stretch values.  These trends are significant at the $10\sigma$ level for $c$ and $5\sigma$ for $x_1$ in the data.  

In previous analyses like JLA or PS1, Hubble residual biases were corrected only as a function of redshift.  Here we find that bias corrections also depend on $c$ and $x_1$.  To illustrate the improvement with a more proper treatment of bias corrections, we measure the distance bias in bins of $\Delta z=0.015*(1+2\times z)$, $\Delta c =0.05$ and $\Delta x_1=0.50$ from simulations and apply these corrections to the data. These corrections remove the trends seen in Fig. 3.  The correction for the distance biases presented here is more accurate than a simplistic bi-linear fit of Hubble residuals for $c>0$ and $c<0$, as done in S14a and \cite{Rubin15}, because of the non-linearity of the trend and its dependance on redshift.  

We find the improvement from our 3-dimensional bias correction comes from the $z>0.1$ sample of PS1, SDSS and SNLS SNe and the $z<0.1$ Low-z sample separately.   A $\sigma_{\textrm{int}}$ is added to the distance errors of the low-z and high-z sample such that $\chi^2/\textrm{NDOF}=1$ for the Hubble diagram fit.  In total, the $\chi^2$ for our sample of 920 SNe is reduced from $920$, using a correction solely based on $z$, to $810$ using the 3-dimensional corrections from simulations with the G10 model and to $821$ using our simulations with the C11 model.  While technically the correction used here adds 12 more DOF from 3 $x_1$ parameters and 3 $c$ parameters for the low and high-z surveys each, past analyses (e.g. B14) used the same parameters in their simulation to correct their distances.  The difference between this analysis and past ones is not that we are using more information, rather that we are optimally using the same information.  For $z<0.1$, $\sigma_{\textrm{int}}$ is reduced slightly from $\sigma_{\textrm{int}}=0.130$ to $\sigma_{\textrm{int}}=0.128$ after applying the correction with the G10 model.  However, for $z>0.1$,  $\sigma_{\textrm{int}}$ is reduced from $0.101$ to $0.083$.  The total dispersion of the SN distances relative to a best fit cosmology is reduced from $0.142$ to $0.130$ in the high-z sample and $0.147$ to $0.142$ in the low-z sample.  These results are very similar to that seen in our simulations.  For the low-z simulated sample, the difference in $\sigma_{\textrm{int}}$ is marginal ($\sim 0.002$) but for the high-z simulated sample, $\sigma_{\textrm{int}}=0.102$ before the correction and $\sigma_{\textrm{int}}=0.084$ after the correction.  The difference between the impact of the correction at high and low-z can be explained by the differences in width of the underlying distributions shown in Table 1.  Given narrower underlying populations for the higher-z samples, combined with larger statistical uncertainties, a larger fraction of the fitted $c$ and $x_1$ values lie outside the range of values that correlate with luminosity and therefore the correction has a larger impact compared to the low-z sample.   
\begin{figure}
\centering
\epsscale{1.15}  
\plotone{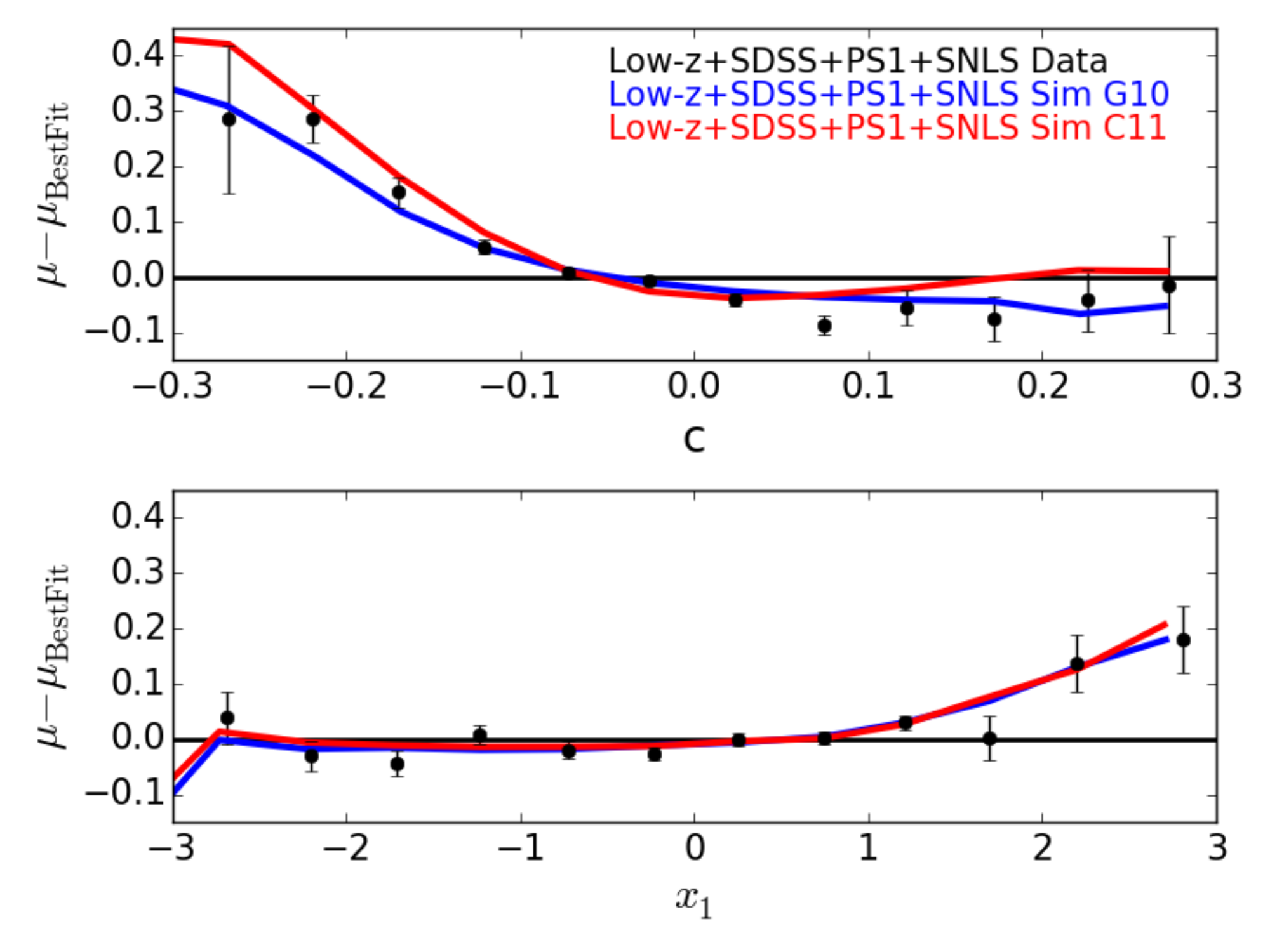}
\caption{The trends in Hubble residuals, $\mu-\mu_{\textrm{BestFit}}$ expressed as the differences between the distances and a best fit cosmology versus $c$ (top) and versus $x_1$ (bottom) for data (points) and simulations (curves).}
\label{fig:filters}
\end{figure}

We show in Fig. 4 that $\sigma_{\textrm{int}}$ still has a strong dependance on the color of the SNe even after the distance biases have been removed.  To better probe this trend, only $z>0.1$ SNe (SDSS+SNLS+PS1) and bins with $>10$ events are shown.  The simulated predictions of $\sigma_{\textrm{int}}$ versus color show marginal agreement with the data in Fig. 4, although the C11 model has better agreement for redder SNe. Neither model predicts the small $\sigma_{\textrm{int}} = 0.0$ at $c \sim -0.12$ (bin with 85 SNe).

Our results are based on simulations with a fixed input value of $\beta$, and thus we try to predict the $\sigma_{\textrm{int}}$ variation in Fig. 4 by varying the simulated $\beta$ using a Gaussian distribution with $\sigma_{\beta}=1$ over a range of $1<\beta<5$.  However, this $\beta$ variation does not reproduce the sharpness of the effect from blue to red colors seen in the data. The minimum of $\sigma_{\textrm{int}}$ in the top panel of Fig. 4 is around where the amount of dust extinction is negligible ($c\sim-0.1$), so the distances for SNe in this color range are less sensitive to issues with $\beta$.

For comparison, our distance correction reduces the $\chi^2$ of the sample $5\times$ more than the distance correction for the step between Hubble residual and host galaxy mass (e.g. \citealp{Sullivan10}, \citealp{Kelly2010}, \citealp{Lampeitl2010}).  However, our distance bias correction has only a very small effect on the `step' in Hubble residuals for SNe with host galaxy mass $>10^{10}~\textrm{M}_{\textrm{solar}}$ and $<10^{10}~\textrm{M}_{\textrm{solar}}$, which we find to be $0.060\pm0.009$ mag before our bias correction and $0.050\pm0.009$ mag after our bias correction.  

Finally, we evaluate the impact of our improved Hubble bias corrections on cosmological constraints from combining our SN sample with CMB measurements from \cite{Planck14}. These constraints are roughly the same using a 3-dimensional Hubble bias correction over $z$, $x_1$ and $c$ values versus the conventional 1-dimensional correction over $z$ values.  However, there is a reduction of $0.003$ in the w-uncertainty (from $w_{\textrm{err}}\sim0.05$) when using the 3-dimensional correction because of the reduced dispersion.  We also compare the value of $w$ recovered when correcting the distances using our populations determined in this analysis versus the populations used in the previous analyses.  We measure a small bias of $\Delta w=+0.005$ for the full sample using the JLA population parameters.  We measure a larger bias of  $\Delta w=+0.025$ for the sample used in the PS1 analysis, which only included PS1 and a low-z sample, mainly due to the difference in color population parameters we find for the low-z sample in this analysis compared to S14b.  Bias corrections using inaccurate populations result in a $w$-bias which depends on the particular combination of SNIa samples.

\begin{figure}
\centering
\epsscale{1.15}  
\plotone{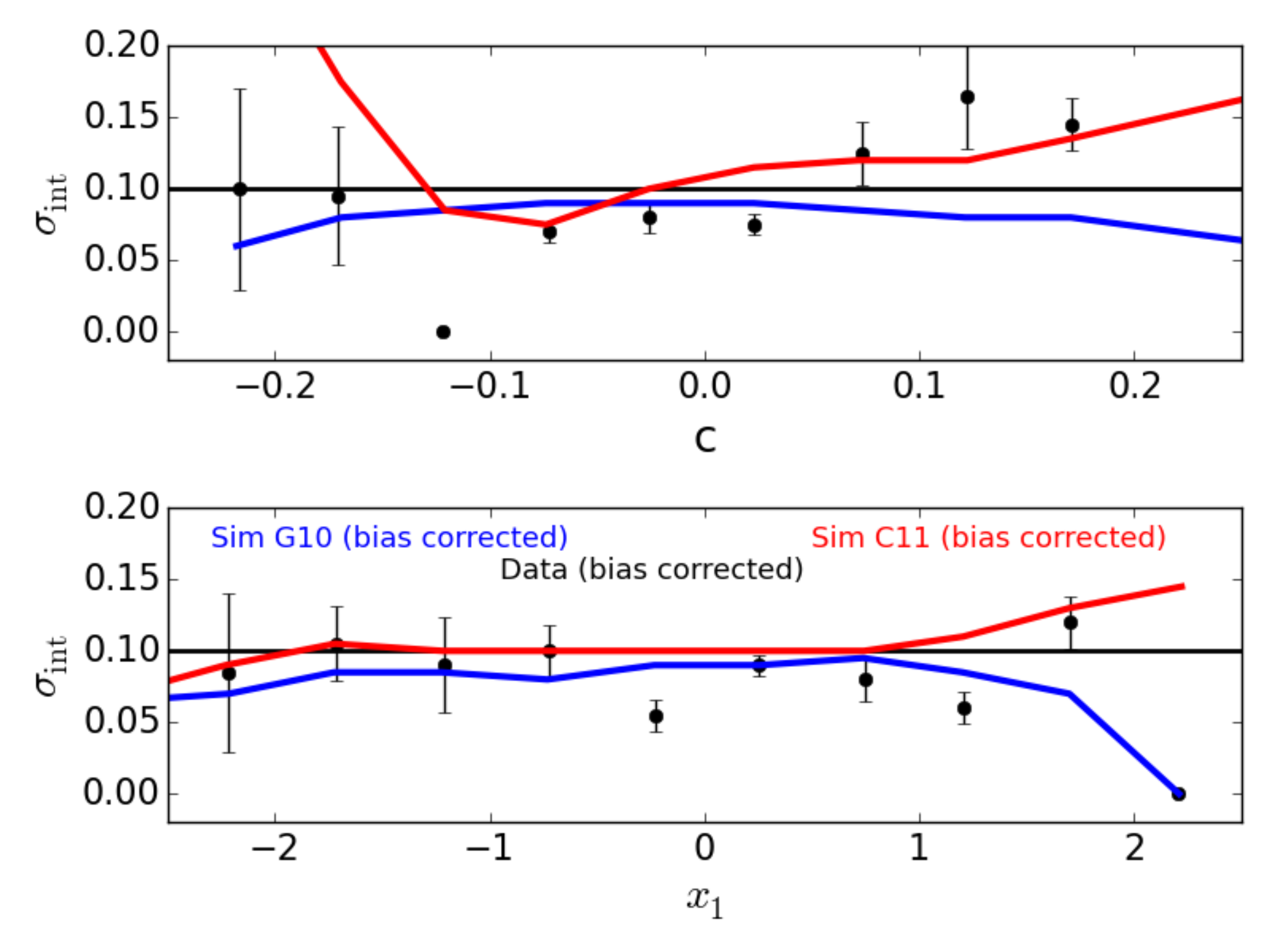}
\caption{The trends in $\sigma_{\textrm{int}}$ of the distances versus $c$ (top) and $x_1$ (bottom) for data and simulations (both for $z>0.1$, once the distance biases shown in Fig. 3 are corrected.  The data shows a significant drop at $c=-0.1$ that is not predicted in the simulation.  Only PS1, SNLS, and SDSS data/simulations are used for this figure.}
\label{fig:filters}
\end{figure}

\section{Discussion}

This paper probes the importance of using simulations to measure the underlying SNIa stretch and color distributions, and shows that a better understanding reduces distance biases and dispersion.  Here, we independently determine the underlying $c$ and $x_1$ distributions, and afterwards, determine the correlation coefficients ($\alpha$ and $\beta$) with luminosity.  In principle, this analysis can be generalized to simultaneously determine the population parameters, $\alpha$ and $\beta$, and the cosmological parameters.  For now, we have verified that the populations found here have little dependance on the assumed cosmology in our simulations.

We have begun to explore additional complexity in our simulations, such as the possible variation in $\beta$.  This variation has already been discussed in previous analyses (e.g. \citealp{Foley11}, \citealp{Burns14}), and should be further explored.  It is particularly interesting that given a large amount of color variation (the C11 scatter model), the underlying color population that we recover resembles a dust model with a sharp cut at the blue color end and an exponential tail at the red color end.  For a Bayesian treatment, one can use our populations as the priors for $x_1$ and $c$ that correlate with luminosity.  However, simulations are also needed to model the selection effects. 

One limitation of our analysis is that we measure the underlying population for only two different models of the intrinsic scatter.  The similarity in the predicted distance biases using the two different scatter models in Fig. 3 shows that the systematic uncertainty on $w$ from the choice of intrinsic scatter model is small.   However, the mismatch between data and simulations shown in Fig. 4 could be a hint of additional unknown biases, so improving the intrinsic scatter model is important for ruling out unknown biases as well as further reducing the Hubble residuals.

In conclusion, this paper presents a robust determination of the underlying populations of the color and stretch populations of SN\,Ia.  These populations are used in simulations to predict and correct for Hubble residual biases, and are essential to infer precise cosmological parameters. 
\section{Acknowledgements}
Analysis was done using the Midway-RCC computing cluster at University of Chicago. This work was generated as part of NASA WFIRST Preparatory Science program 14-WPS14-0048 and is supported in part by the U.S. Department of Energy under Contract DE-AC02-76CH03000.

\bibliographystyle{apj}



\end{document}